\documentclass[aps,pre,twocolumn,groupedaddress]{revtex4-1}

\usepackage{graphicx}
\usepackage{dcolumn}
\usepackage{bm}
\usepackage{amsmath}
\usepackage{xfrac}
\usepackage[table]{xcolor}
\usepackage{mathtools}
\usepackage{changes}

\begin{document}

\title{Symmetry in Critical Random Boolean Network Dynamics}

\author{Shabnam Hossein}
\affiliation{Department of Physics, University of Houston, Houston, Texas 77204-5005, USA\\
Texas Center for Superconductivity, University of Houston, Houston, Texas 77204-5002, USA
}

\author{Matthew D. Reichl}
\altaffiliation{Current address: Laboratory of Atomic and Solid State Physics, Cornell University, Ithaca, New York 14853, USA}
\affiliation{Department of Physics, University of Houston, Houston, Texas 77204-5005, USA\\
Texas Center for Superconductivity, University of Houston, Houston, Texas 77204-5002, USA
}

\author{Kevin E. Bassler}
\email[]{bassler@uh.edu}
\affiliation{Department of Physics, University of Houston, Houston, Texas 77204-5005, USA}
\thanks{Permanent address}
\affiliation{Texas Center for Superconductivity, University of Houston, Houston, Texas 77204-5002, USA}
\affiliation{Max Planck Institute for the Physics of Complex Systems, N\"{o}thnitzer Str. 38, Dresden D-01187, Germany}

\date{\today}

\begin{abstract}
Using Boolean networks as prototypical examples, the role of symmetry
in the dynamics of heterogeneous complex systems is explored. We show that symmetry of the dynamics,
especially in critical states, is a controlling feature
that can be used both to greatly simplify analysis and to characterize different types of dynamics. 
Symmetry in Boolean networks is found by determining
the frequency at which the various Boolean output functions occur.
There are classes of functions that consist of Boolean functions that behave similarly.
These classes are
orbits of the controlling symmetry group. 
We find that the symmetry that controls the critical random Boolean networks is 
expressed through the frequency by which output functions are utilized by nodes 
that remain active on dynamical attractors. 
This symmetry preserves canalization, a form of network robustness. 
We compare it to a different symmetry known to control the dynamics of an evolutionary
process that allows Boolean networks to organize into a critical state. 
Our results demonstrate the usefulness and power of 
using the symmetry of the behavior of the nodes 
to characterize complex network dynamics, and introduce
a novel approach to the analysis of heterogeneous complex systems.
\end{abstract}


\maketitle

\section{Introduction}

The concept of symmetry has played a pivotal role in the advancement of the twentieth century physics. 
Accounting for symmetry can greatly simplify analysis and provide powerful insights into 
the nature and behavior of physical systems.
Group theory is the mathematical framework that translates symmetry into suitable mathematical relations. 
Symmetry and the theory of groups have proven useful in many areas of physics.  
In particle physics, symmetries have become an indispensable tool of theory formation. 
It was the classification of hadrons through representations of symmetry group $SU(3)$ 
that suggested common constituents for these particles and led to the Standard Model~\cite{griffiths_elementary_2005}. 
Likewise, in traditional condensed matter physics, 
symmetry is important for determining the behavior and characteristics of systems,
from the structure of lattices to universal aspects of critical phenomena~\cite{kadanoff_statistical_2000}.

Recently, complex systems and more specifically complex networks have received a lot of 
interest by physicists~\cite{boccara_modeling_2010, albert_statistical_2002, richter_komplexe_2002, claudius_complex_2008}. 
Complex systems often differ from more traditional condensed matter systems because they consist of
heterogeneous components~\cite{boccaletti_complex_2006}, 
which makes their analysis and achieving any sort of
general understanding of their behavior difficult.
A question that naturally arises then is: Can symmetry in heterogeneous 
complex systems be exploited to simplify their analysis 
and obtain fundamental insights into their dynamics? 
In order to answer this question, it is essential to determine the role that symmetry has in 
controlling the behavior of these systems,
and what it can tell us about their structure. It is also important to know whether 
symmetry can be used to distinguish systems with different dynamics.

Boolean networks are an ideal type of complex system with which to answer these questions.
This is because Boolean networks, despite being relatively simple, have essential features 
of complex systems, including heterogeneous structure and dynamics of their constituent 
parts, and display non-trivial dynamics. Most notably, there exists a continuous phase 
transition in their dynamical behavior~\cite{derrida_phase_1986}.
Boolean networks were introduced to model the genetic regulatory systems of 
biological cellular organisms~\cite{kauffman_homeostasis_1969}, 
and have been used to successfully model
essential features of the wild-type 
gene expression patterns of the \textit{Drosophila melanogaster} segment polarity 
genes~\cite{albert_topology_2003}, the regulatory sequences 
of models of \textit{Arabidopsis thaliana} flowers~\cite{espinosa-soto_gene_2004, pandey_boolean_2010}, 
and the cell-cycle network of yeast (\textit{S. cerevisiae})~\cite{li_yeast_2004}. 
They have also 
been used extensively to model other biological systems, 
such as
neural networks~\cite{bornholdt_topological_2000, wang_oscillations_1990}, 
as well as a wide range of physical
and social systems~\cite{aldana_boolean_2003, wang_boolean_2012,bornholdt_less_2005},
including switching circuits~\cite{harrison_introduction_1965}
and social groups within which consensus emerges through peer interactions~\cite{green_emergence_2007}. Boolean networks also count as paradigmatic examples of complex systems~\cite{gong_quantifying_2012}.

In their original variant, which is also considered here,
Boolean networks consist of $N$ nodes with binary output states, 0 or 1, that 
are connected by a directed graph of in-degree $K$ describing their interactions. The output state of the
nodes are updated synchronously and periodically, 
according to the input each node receives from the nodes it is connected to by its in-links, as follows.
The state of node $i$ at time $t$, $s_{i}(t)$, is the output state of a 
fixed Boolean function, $f_{i}$, assigned to node $i$,
\begin{eqnarray*}
s_{i}(t)&=&f_{i}(s_{i_{1}}(t-1), s_{i_{2}}(t-1), \ldots, s_{i_{K}}(t-1) ),
\end{eqnarray*}
where $s_{i_{1}}(t-1), s_{i_{2}}(t-1), \ldots, s_{i_{K}}(t-1)$ are the states  
of the $K$ nodes connected to node $i$ by its in-links, at time $t-1$. 
Note that the Boolean function $f_{i}$, in general, varies for different nodes, which, 
along with the differences in their respective connections in the directed graph,
gives them heterogeneous dynamics.
In \textit{random} Boolean networks, each output function, $f_{i}$, is assigned randomly 
from the set of $2^{2^{K}}$ possible Boolean functions with $K$ inputs, and the out-links 
of the directed interaction graph are also chosen randomly.

Depending on the distribution of their Boolean functions, 
ensemble-averaged random Boolean networks have two distinct phases of 
dynamical behavior: \textit{chaotic} and \textit{frozen}~\cite{derrida_phase_1986}. 
One of the features that distinguishes these two phases is the 
distribution of network attractor periods~\cite{bastolla_numerical_1997}. 
Because time is discrete in the dynamics, and the network's 
state space is both discrete and finite, starting 
from any initial set of node states, the network dynamics
will always eventually settle, in finite time, onto a periodic attractor. 
In the chaotic phase, the attractor distribution is sharply peaked around 
an average period that grows exponentially with the network size. 
In the frozen phase, on the other hand, the distribution of attractor periods 
is independent of the network size. 
Critical networks, at the boundary between the two phases, 
have a broad, power law distribution of attractor periods~\cite{bastolla_numerical_1997, bhattacharjya_power-law_1996}. 
Power law distributions are scale free.
Scale free behavior is characteristic of all condensed matter 
systems at continuous phase transitions, at which there are no characteristic 
lengths or time scales~\cite{yeomans_statistical_1992}. As a result, the symmetry properties of the system are particularly important at the critical state - unlike other phases 
of dynamics where characteristic periods describe the behavior of the system.

Critical networks are of particular interest in applications, 
since, for example, they can store and transfer more information than 
frozen networks with static outputs or chaotic networks with 
random outputs~\cite{packard_adaptation_1988}. 
They can be obtained either by choosing a critical distribution of Boolean functions
by construction, or by implementing a process according to which the distribution of
Boolean functions evolves. 
A Boolean function is defined by its output table, which lists its Boolean output value 
for each of the $2^K$ possible values of $K$ distinct Boolean inputs. It can be constructed with a bias towards homogeneous output
by independently choosing each of the elements in its output table to 
be $1$ with probability $p$, or $0$ with probability $1-p$.
If $p\!=\!1/2$, then all the different functions are equally likely to be constructed,
but if $p\neq1/2$, there is a bias in favor of constructing
functions whose outputs are more homogeneous.
Critical random Boolean networks can be constructed by randomly choosing Boolean functions 
with a critical bias, $p_{c}$. 
The critical bias $p_{c}$ is the value that satisfies the equation
$1\!=\!2Kp_{c}(1-p_{c})$~\cite{derrida_random_1986}. 
Critical Boolean networks can also be obtained, 
for example, through an 
evolutionary game, played by the nodes, that selects for diversity in their behavior~\cite{Paczuski_self-organized_2000}.

In this paper, we explore the role of symmetry in the dynamics of Boolean network models
of complex systems. In particular, we consider symmetry in the behavior of the nodes. 
This approach contrasts with studies of complex networks that have considered symmetry in the 
topological structure of the links, such as Ref.~\cite{macarthur_spectral_2009}.
By analogy with other condensed matter systems, it is expected that the importance
of symmetry will be most evident in critical networks, at the continuous transition,
``at the edge of chaos''~\cite{kauffman_origins_1993}.
We first find how symmetry manifests itself in Boolean networks, and find a symmetry of the critical state of constructed random Boolean networks. 
For this, 
we look at the frequency distribution of the heterogeneous Boolean functions of the nodes
that remain active on an attractor in critical networks, and find the ratios that depict what fraction of nodes with a particular Boolean function remain active on a dynamical attractor.
We show that there are classes of functions, related by symmetry, that occur with the same ratio. We then determine the nature of this symmetry and show that
it is related to preserving a form of network robustness known as \textit{canalization}, which is 
of importance in developmental biology~\cite{waddington_canalization_1942}.
Canalization preserving
symmetry is also compared and contrasted with that of the dynamics of the evolutionary game 
mentioned above in order to show how different dynamics at the 
critical state correspond to different symmetries. 
Finally, we discuss how our results demonstrate the usefulness and power of 
using symmetry for characterizing complex network dynamics, and point to potential
new approaches to the analysis of complex systems.


\section{Symmetry in Critical Dynamics}

As stated earlier, heterogeneous composition is often a
feature of complex systems that distinguishes them from more traditional
condensed matter systems. This
heterogeneity makes them complicated, but, as we will now see, also allows
for novel methods of analysis that can provide a fundamental understanding 
of their behavior. In particular, one can look for symmetry among the
heterogeneous components of a complex system to find the nature of its dynamics.

In Boolean networks, one aspect of their heterogeneity is that nodes
can behave differently for identical sets of inputs. The responses of the 
nodes are dictated by their output functions,
which are heterogeneously distributed.
Moreover, various
Boolean functions are generally utilized with different frequencies, and nodes
with different Boolean functions can play different roles in the dynamics
of a network.
To look for a manifestation of symmetry in the dynamics in the
critical state of random Boolean networks, consider the distribution of Boolean functions
utilized by the nodes.

We have simulated ensembles of 
networks constructed to be critical, by randomly forming the function of each node with critical bias while maintaining the parity between 0 and 1. 
 This parity is achieved obtained by randomly choosing to bias the output of each node's function toward a homogeneous output of either 0 or 1, with equal probability.
This form of bias toward homogeneity of the output functions is used, for
simplicity, to maintain an overall symmetry between $0$s and $1$s in all
aspects of the networks' structure and dynamics.
Also, all networks that we consider here have simple directed graphs describing the interactions
of the nodes; that is, the nodes have no self-interactions and receive at most one input
from any other node.
Furthermore, our emphasis in this article is on networks with $K\!=\!2$ and $K\!=\!3$, although analogous observations and arguments, with similar results expected, can be made for networks
of nodes with any in-degree $K$.

\begin{figure}
\begin{center}
\includegraphics[width=8.6cm,keepaspectratio]{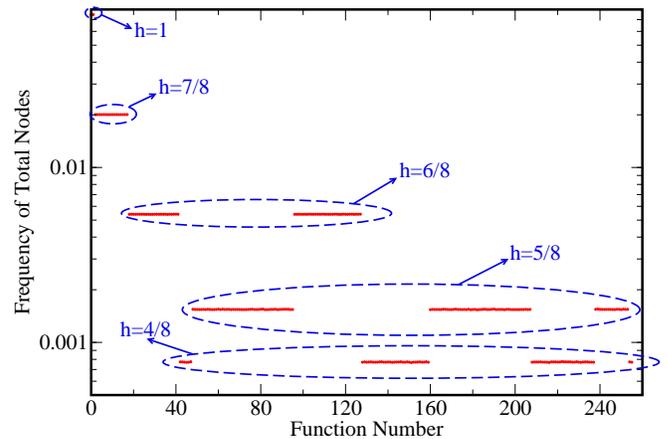}\noindent \\
\end{center}
\vspace{-0.5cm}
\caption{Ensemble-averaged frequency of the Boolean functions
used by all nodes 
of critical random Boolean networks with in-degree $K\!=\!3$. 
Functions that have the same frequency have the same homogeneity value. 
Five different homogeneity values $h$ are possible, as shown. The frequencies are normalized such that the sum of the frequencies of all of the functions is unity.}
\label{FigFreqk3_1}
\end{figure}

\begin{figure}
\begin{center}
\includegraphics[width=8.6cm,keepaspectratio]{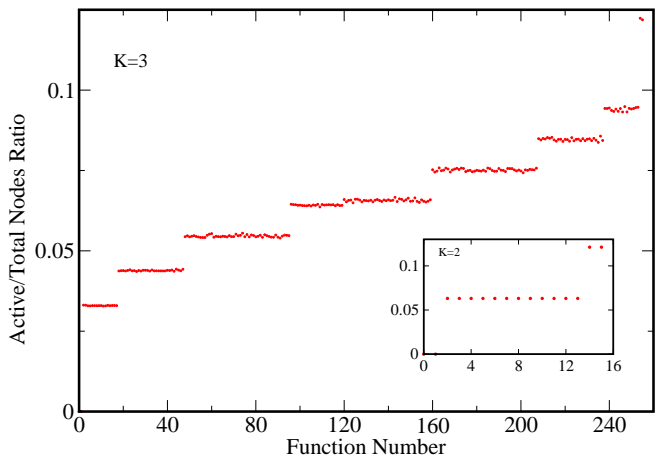}\noindent 
\end{center}
\vspace{-0.5cm}
\caption{
Ratio of the ensemble-averaged number of the Boolean functions
used by nodes that are active on dynamical attractors to that used by all nodes 
of critical random Boolean networks with in-degree $K\!=\!3$. 
Ten distinct classes of functions that occur with the same ratio can be recognized. 
The inset shows analogous results for networks with in-degree $K\!=\!2$, 
where 3 function classes are found. It should be noted that the first class in both $K\!=\!2$ and $K\!=\!3$ cases consists of the two constant Boolean functions, numbered 0 and 1. Obviously the ratio is 0 for this class, since the nodes using these functions are frozen from the beginning. The statistical errors are small compared to the symbol size.}
\label{FigFreqk3_2}
\end{figure}

Figures~\ref{FigFreqk3_1} and \ref{FigFreqk3_2} 
show numerical results for networks 
of $10^5$ nodes each with $K\!=\!3$ inputs, averaged over an ensemble of $10^6$ networks.
There are $256$ possible Boolean functions with $3$ inputs, each of which occurs with a certain frequency. 
In these figures, the functions are numbered from 0 to 255, in a manner that groups 
related functions together~\cite{liu_finite_2011}. The numbering of the functions
is the same in both figures. 

The distribution of Boolean functions among all nodes of the network
is shown in Fig.~\ref{FigFreqk3_1}.
Five different frequencies are possible due to the binomial distribution of the 
fraction of $1$s in functions' output tables, and the imposed symmetry of $0$s
and $1$s in the network.
The characteristic of a function that determines its frequency in this case
is its internal homogeneity. 
Internal homogeneity, which is a measure of how homogeneous the output of the function is, is defined as
the fraction of the $2^K$ outputs of a function that are $0$ or $1$, 
whichever is larger~\cite{walker_temporal_1966}.
Thus for Boolean functions with $3$ inputs, there are $2^{3}\!=\!8$ possible input sets, 
and the homogeneity of functions can take 
five values: $1$, $\sfrac{7}{8}$, $\sfrac{6}{8}$, $\sfrac{5}{8}$, and $\sfrac{4}{8}$. 
If the homogeneity of a function is $h$, 
then its frequency is
\begin{eqnarray*}
\text{Freq}(h)&=&\frac{1}{2}\binom{8}{8h}\left[
p_{c}^{8h}(1-p_{c})^{8-8h}
+
p_{c}^{8-8h}(1-p_{c})^{8h}
\right].
\end{eqnarray*}

Not all the nodes, however, have the same importance for the dynamics of 
the network. Some nodes never change their output states, 
either because they have completely homogeneous
output functions, or because they receive input from functions with homogeneous
output functions.
These 
\textit{frozen} nodes are irrelevent to the dynamics of the network~\cite{bastolla_relevant_1998, socolar_scaling_2003, kaufman_relevant_2006, samuelsson_exhaustive_2006}. 
This suggests that in order to find symmetry in the dynamics of the network, 
analysis should focus on the behavior of the non-frozen, \textit{active} nodes of the network 
and on their functions. 

Figure~\ref{FigFreqk3_2} shows the ratio of the number of Boolean functions occurring on
nodes that are active on dynamical attractors
to the number at which they occur on all nodes. 
Investigating this ratio, rather than the frequency of the Boolean functions 
on active nodes, allows the effect of the initial conditions to be eliminated, 
and the functions to be properly compared.
Remarkably, only 10 different ratios are realized.
The functions that have the same ratio form a \textit{class}.
With the function numbering that has been chosen, all functions in a class are grouped
together, and the classes are
ordered from the lowest to the highest ratio. 
The inset of Fig.~\ref{FigFreqk3_2} shows analogous results for 
critical random Boolean networks with $K\!=\!2$ inputs per node. 
In this case, 3 different function classes occur.

Functions have the same ratio and are, therefore, in the same
class because they behave similarly.
This indicates that 
there is a symmetry between the functions in the same class. Moreover, 
since the active nodes determine the dynamics of the network, 
this symmetry must be a characteristic of the dynamics of the network. Thus, the 10 classes of functions for networks with $K\!=\!3$, 
and the 3 classes for networks with in-degree $2$ are exactly 
the manifestations of the symmetry that we were looking for in order to characterize the dynamics. 
In principle, different dynamics can have different symmetries, 
and they can be distinguished
by the behavior of the heterogeneous parts of the networks.

\section{Analytic Description of the Dynamics}
\label{sec:III}

In order to analytically describe the numerical observations from the previous section, 
it is important to understand how
the active nodes were identified. 
This was done using a \textit{pruning} method~\cite{kaufman_scaling_2005}. 
In this method, the frozen nodes are found starting from the ones that have homogeneous output functions, which are by definition frozen. Other frozen nodes are found given the fact that 
nodes with frozen outputs can cause nodes
that receive inputs from them to freeze; even if those nodes have
inhomogeneous output functions, which, in turn, can cause even more nodes to freeze. After finding all the frozen nodes, 
the active nodes are identified as those that remain.

A stochastic model of this pruning process 
was used to analytically
calculate the scaling properties of the
number of active nodes in the critical state of Boolean networks~\cite{kaufman_scaling_2005, mihaljev_scaling_2006}.
Here we use a variant of this stochastic pruning model. Unlike the original variant of the model, 
which tracked only the number of active nodes through the pruning process,
we track the independent progress of nodes with various output functions. 
This is essential for our purposes because at the end, 
we want to determine the distribution of output functions of the active nodes, and not merely the number of active nodes.

For the sake of clarity, 
we will explain the pruning model for 
networks of nodes with $K\!=\!2$ inputs. 
An analogous calculation, however, can be applied to networks with 
nodes of any in-degree $K$.
Before we define the pruning process, however, let us first consider 
some representations of Boolean functions.
The Boolean functions with $K$ inputs can be ordered by taking their binary outputs
from the $2^K$ possible sets of inputs as a single binary number and then, by converting
that number to a decimal number, such as what is shown 
in Fig.~\ref{Figfunction3table}.
(This is not the decimal numbering of functions used in Figs.~\ref{FigFreqk3_1} and \ref{FigFreqk3_2}.)
They can also be represented geometrically as $K$-dimensional
Ising hypercubes~\cite{reichhardt_canalization_2007}.
In this representation, each of the corners of a hypercube corresponds to
a set of binary inputs, and the color of the corner indicates the corresponding output
value. There is a one-to-one mapping from the set of Boolean functions with $K$ inputs
to the colorings of the $K$ dimensional Ising hypercube.

\begin{figure}[t]\begin{center}
\includegraphics[width=8.6cm,keepaspectratio]{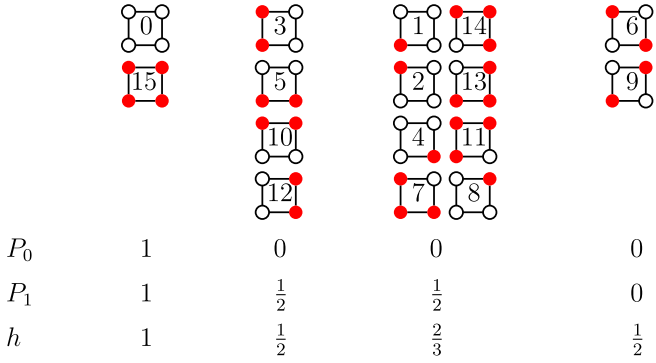}
\end{center}
\vspace{-0.5cm}
\caption{Ising hypercube representation of the $16$ Boolean functions with in-degree $K\!=\!2$. 
The decimal numbering of each function is given inside its corresponding Ising square representation.
The canalization values, $P_{k}$s, and the internal homogeneity, $h$, of the functions are also shown. 
In a hypercube representation, $P_{k}$ is the fraction of $K\!-\!k$ dimensional 
hypersurfaces of the hypercube that are homogeneously colored. 
In this case, since $K\!=\!2$, $P_0$ is 1 if the function is a frozen function, represented by
a homogeneously colored square, and is 0 otherwise. 
$P_1$ is the fraction of the homogeneously colored edges in the representation of the function. 
Functions with the same set of $P_k$s and $h$ are grouped together.}
\label{Figk2functions}
\end{figure}

The Boolean functions with $K\!=\!2$ inputs, for instance,
can be numbered in this way from 0 to 15. As shown in Fig.~\ref{Figk2functions}, they can also be represented geometrically as colorings of the Ising square.
In the geometric representation, 
assuming that each edge of the square is of unit length, 
if the square is placed on an $x-y$ plane with 
its lower left corner at the origin, the coordinates
of each corner map to its corresponding input set.
Thus, the horizontal or $x$ coordinate corresponds to the first input value,
while the vertical or $y$ coordinate corresponds to the second input value.
White and red corner colorings indicate outputs of 0 and 1, respectively.

We begin with a mean-field version of the pruning model.
This model assumes that each node has an assigned output function, but
the network links are treated probabilistically.
Consider a collection of urns, each of which
contains nodes of a particular type. 
There is a separate urn for nodes with each different type of Boolean function, which we call \textit{original} urns, and an urn for frozen nodes. There are also urns for partially pruned nodes, which 
we call \textit{intermediate} urns. 
Throughout the pruning process, we track the number
of nodes in each urn, 
which is a continuous variable in the model.
Begin by placing every node in the urn corresponding to the type of output function
it has. Let $N_i$ be the number of nodes in the original urn containing
nodes of type $i$. Initially, $N_i$ is the
total number of nodes that have function $i$,
and the total number of nodes in all urns is $N\!=\!\sum_{i=0}^{15} N_i$, which is the total number of nodes of the network. 
Then move all nodes with homogeneous output functions, 
that is all nodes with functions 0 and 15,
into the frozen urn; so that now $N_0\!=\!N_{15}\!=\!0$
and $N_F\!=\!N_0+N_{15}$.

The pruning process proceeds step by step
by picking a random node from the frozen urn
and considering its probabilistic influence on other nodes. 
That is, the probability that each remaining non-frozen node
will have an in-link from the chosen node, therefore receives
frozen input from it, is calculated.
These probabilities are used to change the population
of nodes in the urns.
Depending on the output function of a node
receiving a frozen input, one of two things
can happen: either the node itself will become frozen, or
it will remain active, but will effectively have an output function
with one less input.
In the latter case, the node becomes partially pruned. 
In either case, the population of the original urn will decease,
and either the population of the frozen urn, or the population of
the appropriate intermediate urn, will increase. 
Thus in each time step of the pruning process,
the probability of the above events are calculated for all the nodes of the network, the population of all urns are updated, and the frozen node is discarded from the frozen urn. 

The process continues either until the frozen urn is empty, 
or until all the nodes of the network become frozen. 
It should be noted that throughout the process, the nodes' functions do not actually 
transform to another function, but they effectively behave as if they have been transformed.

\begin{figure}[h!]
\begin{tabular}{ c | c || c }
    input 1 & input 2 & output \\ \hline
    0 & 0 & 1 \\ 
    0 & 1 & 1\\
    1 & 0 & 0 \\
    1 & 1 & 0\\  
  \end{tabular}
\caption{Example of a Boolean function with $K\!=\!2$ inputs.
This is function number 3 because the output has binary number 0011. 
Note that if the first input value is fixed to either 0 or 1, 
the output value is determined 
independent of the value of the second input. 
Thus, both values of the first input are canalizing.}
\label{Figfunction3table}
\end{figure}

\begin{figure}
\begin{center}
\includegraphics[width=7.0cm,keepaspectratio]{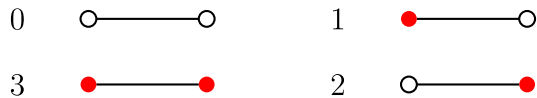}\noindent\\ 
\end{center}
\vspace{-0.5cm}
\caption{Ising hypercube representation of the four Boolean functions with in-degree K\!=\!1. There are two constant functions (0 and 3), and two non-constant functions (1 and 2). The non-constant functions are related by parity and make the function pair $K1$.}
\label{K1function}
\end{figure}

Although one can construct a set of equations that describes the dynamics 
of the populations of all the urns listed above, the description can be simplified
by accounting for the fact that the critical dynamics as constructed here maintains 
a symmetry between the output values, 0 and 1.
Therefore, all original and intermediate urns containing functions related by parity behave identically. 
The parity of a function is obtained by changing all the output values of the function,
that is, swapping every value of 1 for 0 and vise versa. 
Because of this observation,
functions related by parity can be placed in the same urn,
and so, instead of 16 original urns, 
a reduced set of 8 original urns can be considered.
This reduced set of original urns contains functions within the pairings
\{0,15\}, \{1,14\}, \{2, 13\}, \{3, 12\}, \{4,11\}, \{5, 10\}, \{6, 9\}, and \{7,8\}. 
Furthermore, in this simplified description, because the only two non-constant Boolean functions
with $K\!=\!1$ input are related by parity, there will be only one function pairing, denoted by $K1$ in the intermediate urns labeling, that correspond to the effective $K\!=\!1$ function behavior of partially pruned nodes.
Figure \ref{K1function} shows the geometric representation of all $4$ possible Boolean functions with in-degree $K=1$. 
In what follows, urns will be designated 
by the primed number of the lowest numbered Boolean function of the nodes they contain.

Consider the changes that occur to the population of the urns that 
contain nodes that have functions in the pairing
\{3, 12\}. 
There are three different urns that can contain such nodes. First, the original urn
that has a population of $N_{3^{\prime}}$. Second, the intermediate urn that contains still-active,
partially pruned nodes, with a population of $N_{3^{\prime}K1}$, that are now
effectively utilizing functions with $K\!=\!1$ input. Third, the frozen urn that has a population of $N_F$.

The following equations describe the changes that occur to the population of these urns 
during one step of the pruning process: 
\begin{eqnarray}
&\begin{array}{lcl}
\Delta N_{3^{\prime}}&=&\displaystyle -2\frac{N_{3^{\prime}}}{N},\\
\Delta N_{3^{\prime}K1}&=&\displaystyle \frac{N_{3^{\prime}}}{N} -\frac{N_{3^{\prime}K1}}{N},\\
\Delta N_{F}&=&\displaystyle \ldots+\frac{N_{3^{\prime}}}{N} +\frac{N_{3^{\prime}K1}}{N}+\ldots .\\
\end{array}
\label{N3eqns}
\end{eqnarray}
The first of these equations describes the fact that
a randomly chosen frozen node will provide an input to one of the nodes in the original urn
containing nodes with either function $3$, or $12$, with probability 
$\frac{2}{N}$,
which causes them to be removed from the urn at that rate.
This is because these functions 
have two possible in-links through which to receive input from the frozen node. There are two equally likely possible outcomes for the nodes that 
are removed from this urn.
The first possibility is that
the input from the frozen node is the first input of a node with function $3$ or $12$, 
which is the $x$ value in the geometric representation. 
In this case, 
regardless of whether the frozen input has value 0, or 1,
the function becomes frozen,
because the other input is irrelevent to the output, and so the node is transferred to the frozen urn. 
The other possibility is that
the input from the frozen node is the second input, 
which corresponds to the $y$ value in the geometric representation. 
In this case, 
again regardless of whether the frozen input has value 0, or 1,
the function does not become frozen.
Instead, it remains active with what is effectively a $K\!=\!1$ output function,
and so is transferred to the appropriate intermediate urn, $3^{\prime}K1$.
Nodes with either function $3$, or $12$ that have already been partially pruned, and
are in the intermediate urn can also receive input from the chosen
node through their one remaining free in-link. This occurs with 
probability $\frac{1}{N}$, and when it does, it freezes the output
of the node, requiring the node to be transferred from the intermediate
urn to the frozen urn. 

A set of equations identical to Eqns.~\ref{N3eqns} 
describes the changes in the population of urns that 
contain nodes that have all of the other function pairings except function pairings $0^{\prime}$ and $6^{\prime}$.
The function pairing $0^{\prime}$ has no changes because the nodes with those functions are frozen by definition.
The analogous set of equations that 
describe the changes in the population of the urns that 
contain nodes with functions in the $6^{\prime}$ pairing is  
\begin{eqnarray}
&\begin{array}{lcl}
\Delta N_{6^{\prime}}&=&\displaystyle-2\frac{N_{6^{\prime}}}{N},\\
\Delta N_{6^{\prime}K1}&=&\displaystyle 2\frac{N_{6^{\prime}}}{N} -\frac{N_{6^{\prime}K1}}{N},\\
\Delta N_{F}&=&\displaystyle \ldots+\frac{N_{6^{\prime}K1}}{N}+\ldots ,\\
\end{array}
\label{N6eqns}
\end{eqnarray}
where $N_{6^{\prime}}$ is the population of the original urn and
$N_{6^{\prime}K1}$ is the population of the intermediate urn.
These equations differ from those of functions $3$
and 12, for instance, because in this case, if either input becomes frozen, regardless of its value, the output of the function remains
active, having then effectively a $K\!=\!1$ output function.
Thus, all partially pruned nodes move from the original urn
to the intermediate urn, with a probability of $\frac{2}{N}$.
No node moves directly from the original urn to the frozen urn.
As before, partially pruned nodes become frozen 
with a probability of $\frac{1}{N}$. 
Therefore, complete changes of the population of the frozen urn, and of the total population
in all urns in one step is 
\begin{eqnarray}
&\begin{array}{lcl}
\Delta N_{F}&=&\displaystyle -1
\quad+\quad \sum_{\mathclap{i\in\{1^{\prime},2^{\prime},\ldots ,7^{\prime}\}\backslash6^{\prime}}} \! \frac{N_{i}}{N}
\quad+\quad \sum_{\mathclap{j\in\{1^{\prime},2^{\prime},\ldots ,7^{\prime}\}}} \frac{N_{jK1}}{N},\\
\Delta N&=& \displaystyle-1. 
\end{array}
\label{NFeqns}
\end{eqnarray}
The $-1$ is in both equations because, after each step, the chosen frozen node is discarded, 
and so both the number of frozen nodes and 
the total number of nodes still being considered decreases by 1.

Dividing the equation of the dynamics of the population of
each urn by the last equation for $\Delta N$,
taking the large $N$ continuum limit, 
gives the following set of mean-field deterministic differential equations: 

\begin{align}
&\displaystyle\frac{{\rm d}N_{i}}{{\rm d}N}= \displaystyle 2\frac{N_{i}}{N}   && i\in \{1^{\prime},2^{\prime},\ldots,7^{\prime}\}\nonumber\\
&\displaystyle\frac{{\rm d}N_{iK1}}{{\rm d}N}=\displaystyle \frac{N_{iK1}}{N}
-\frac{N_{i}}{N}
&& i\in \{1^{\prime},2^{\prime},\ldots,7^{\prime}\}\backslash 6^{\prime}\nonumber\\
&\displaystyle\frac{{\rm d}N_{6^{\prime}K1}}{{\rm d}N}=\displaystyle 2\frac{N_{6^{\prime}K1}}{N}
-\frac{N_{6^{\prime}}}{N}\label{finaleqn} \\
&\displaystyle\frac{{\rm d}N_{F}}{{\rm d}N}=\displaystyle \frac{N_{F}}{N} 
-\frac{N_{6^{\prime}}}{N}.\nonumber
\end{align}

This set of coupled equations can be solved analytically. 
The condition $N_{F}\!=\!0$ determines the end of the pruning process. 
When this condition is met, the sum of the populations of all urns gives 
the total number of active nodes. 
However, more detailed information can also be discerned.
In particular, the number of the active nodes with, say, function $3$ 
can be found by summing 
the population of the original and the intermediate urns for nodes with function 
pairing $3^{\prime}$ and dividing it by 2, $\frac{1}{2} (N_{3^{\prime}}+N_{3^{\prime}K1})$. 
Thus in this way, this variant of the pruning model allows for two quantities to be calculated:
population of active nodes with particular output functions, and thereby the frequency of use of different Boolean functions among active nodes.

\begin{figure}[h!]\begin{center}
\includegraphics[width=8.6cm,keepaspectratio]{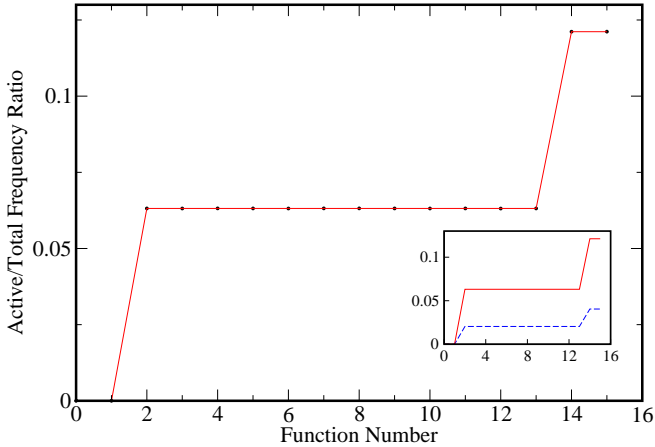}
\end{center}
\vspace{-0.5cm}
\caption{Comparison of the analytical calculations and simulation results for networks consisting of nodes
with in-degree $K\!=\!2$. 
In the main figure, the black circles show simulation results that were also shown in the inset of Fig.~\ref{FigFreqk3_2},
and the red line shows the results of analytical 
calculations described in section~\ref{sec:III}. 
The inset compares the analytical results with (red solid line) and without (blue dashed line) including the 
effect of fluctuations.}
\label{Figk2analytical}
\end{figure}

\begin{figure}[t]\begin{center}
\includegraphics[width=8.6cm,keepaspectratio]{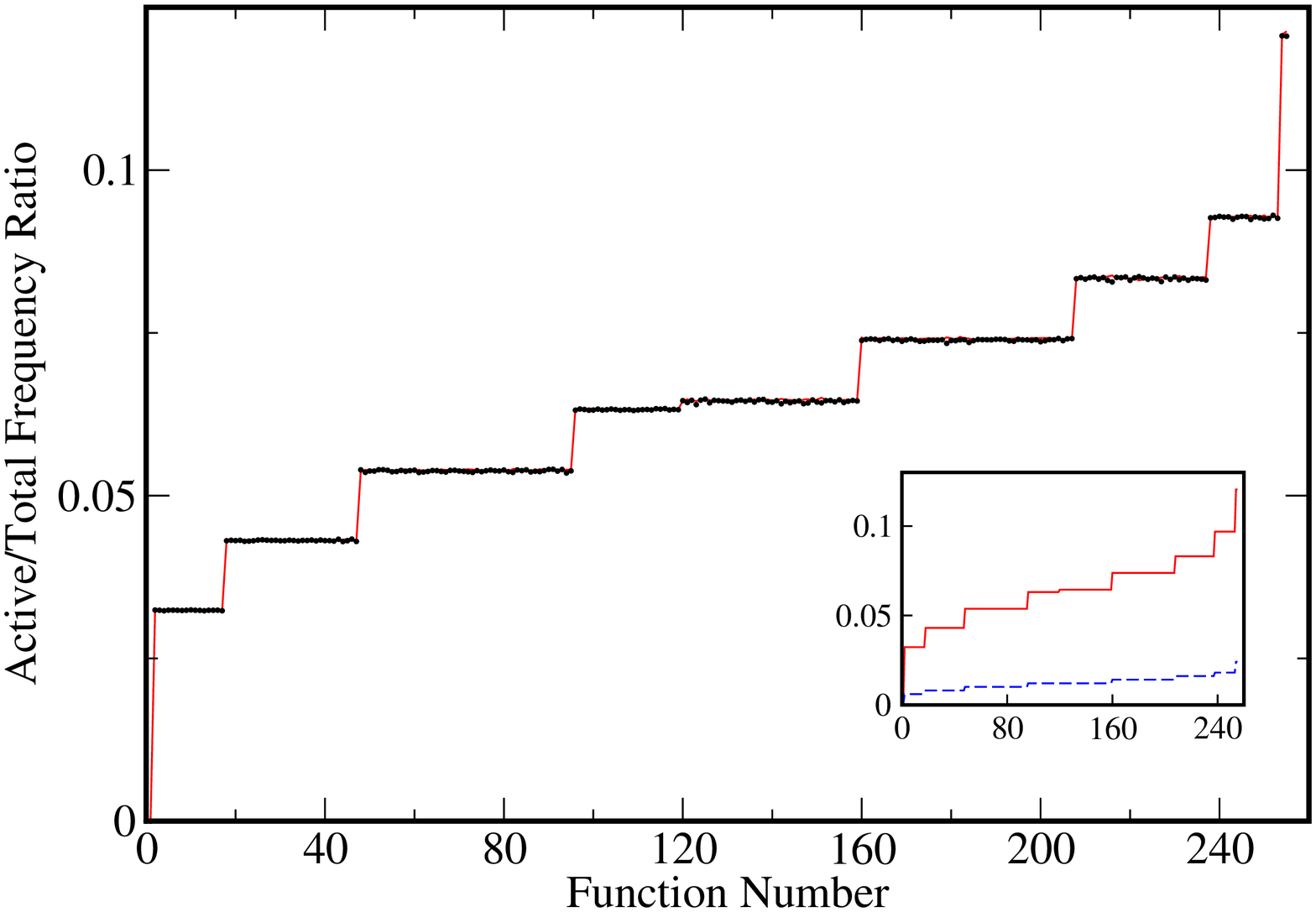}
\end{center}
\vspace{-0.5cm}
\caption{Comparison of the analytical calculations and simulation results for networks consisting of nodes
with in-degree $K\!=\!3$. 
In the main figure, the black circles show simulation results that were also shown in Fig.~\ref{FigFreqk3_2},
and the red line shows the results of analytical 
calculations.
The inset compares the analytical results with (red solid line) and without (blue dashed line) including the 
effect of fluctuations.}
\label{Figk3analytical}
\end{figure}

Unfortunately, this mean-field model yields poor results when describing 
a critical network. This is because, as is well known from condensed matter
physics~\cite{schwabl_statistical_2006}, fluctuations are important in critical systems,
but they are ignored in the model.
Therefore, in order to get more accurate results, 
fluctuations must be taken into consideration. This can be done by adding appropriate noise terms to the differential equations, 
producing a stochastic pruning model,
as was done previously for the earlier variant of the pruning model~\cite{kaufman_scaling_2005}. 
Following previous work on that variant, 
a Fokker-Planck equation corresponding to the stochastic dynamics can be derived and
solved numerically. 
This produces accurate predictions of the active node ratio of the different functions,
as shown in Figs.~\ref{Figk2analytical} and \ref{Figk3analytical} 
for networks with nodes with $K\!=\!2$ and $K\!=\!3$, respectively. 
Note that the insets of these figures, which show the difference between the 
mean-field and the stochastic models, clearly demonstrate the importance of including the effects of fluctuations.
The results from the stochastic model, though, accurately predict the simulation results from
the previous section.

It is important to note that although we have obtained our numerical results, shown in Figs.~\ref{FigFreqk3_1} and \ref{FigFreqk3_2},
through an implementation of the pruning process, 
one could simply naively look for the active nodes of a network 
by following the output states of the nodes on the network's attractors. 
Critical random Boolean networks can have multiple attractors~\cite{samuelsson_superpolynomial_2003, bastolla_modular_1998, bastolla_numerical_1997, bilke_stability_2001, socolar_scaling_2003}. 
The active nodes are the ones whose output values change over the course of some attractor.
Hence in principle, one should examine the behavior of the nodes on all of the attractors to
find the active ones.
If this is done, an identical number of function classes should be detected through such a naive approach. 
On some attractors though, otherwise active nodes may become frozen through dynamical processes
not associated with pruning~\cite{moller_formation_2012}, for example, by forming so-called
self-freezing loops~\cite{paul_properties_2006}. 
Therefore, the total number of active nodes found by using the naive approach may still be less than that found with
a pruning process.
Under particular conditions concerning the initial distribution of Boolean functions among
the nodes, 
which are not met in the case we have considered here,
self-freezing loops can be an important contributor to the frequency of active nodes.
For all other initial function distributions, the contribution from self-freezing loops
vanishes in the large network limit~\cite{mihaljev_scaling_2006}. 
In this case, the difference between using a pruning process and using the naive approach
is negligible, 
and the vast majority of the frozen nodes are found by the iterative pruning process.

\section{Nature of Symmetry of the Critical Dynamics}

Both the simulation and the analytic results show that there are classes of functions
that produce equivalent, symmetric behavior. Indeed, as we have seen in Eqns.~\ref{finaleqn},
the equations for the urns that contain
nodes with functions that are in the same class are identical, making the symmetry of the output functions explicit. Accounting for this symmetry allows for the calculations
described in the previous  section to become dramatically simplified. 
That is, it is not necessary to have separate urns for nodes with
each different function or even function pairings. Instead,
merely considering a collection of urns containing all nodes 
that have functions in the same class yields identical
results.
So what is the nature of the
symmetry between functions in critical random Boolean networks that
produces this identical behavior of functions in a class? 

From the description of the pruning process in critical random Boolean
networks one can see that functions are in
the same class, 
if they have the same probability of becoming frozen at each stage of the pruning process.
In general, this is true when fixing up through $K-1$ inputs for functions with $K$ inputs.
That is, if for Boolean functions with $K$ inputs 
the quantities
$P_{k}$ are defined for $k\!=\!0, 1, \ldots, K-1$ 
as the probability that the output of the function is 
determined if $k$ randomly chosen inputs are fixed, 
but the other inputs are allowed to vary,
then functions with the same set of $P_k$ values are in the same class.
The function classes of the critical dynamics are uniquely determined by their
set of $P_k$ values.
The number of such classes is $2$, $3$, $10$, and $46$ 
for functions of $K\!=\!1$, $2$, $3$, and $4$ inputs, respectively. 

The ability of a Boolean function to have its output become fully determined, 
so that it becomes frozen, when only a subset of the input values are specified,
is a property known as \textit{canalization}.
Subsets of inputs that control the output of a Boolean function in this way are called 
canalizing inputs.
Canalization is a form of network robustness, 
in which the dynamics of the network are insensitive to changes in the non-canalizing
inputs and their connections.
Long recognized as a property of 
developmental biological systems~\cite{waddington_canalization_1942},
canalization is thought to be important because it allows greater evolutionary
variation, hidden behind canalized traits, without potentially deleterious effects~\cite{wagner_robustness_2007}.
How canalizing a Boolean function is can be quantified by its set of $P_k$ values.
Figure~\ref{Figk2functions} shows the canalization values of different functions with $K\!=\!2$ inputs. 

The symmetry of the critical state of random Boolean networks, then, is that
which preserves canalization. Having the same set of $P_k$ values defines an equivalence relation
between Boolean functions with the same number of inputs, and a set of functions with
the same set of $P_k$ values form an equivalence class.
The group that describes the canalization preserving symmetry 
has these equivalence classes.
These equivalence classes are also orbits of that symmetry group;
the functions in a class map into each other through symmetry transformations
that are elements of the group.

\begin{figure}\begin{center}
\includegraphics[width=8.6cm,height=10cm,keepaspectratio]{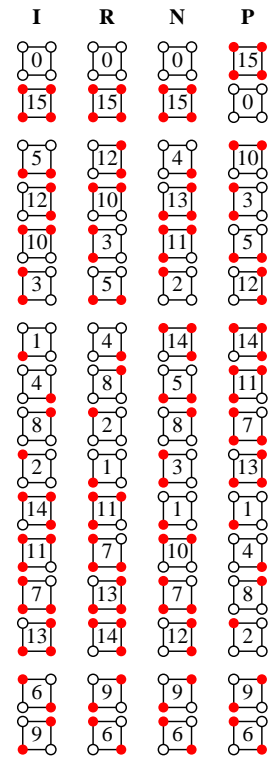}
\end{center}
\vspace{-0.5cm}
\caption{Generators of the minimal canalization preserving symmetry group for Boolean functions with $K\!=\!2$
inputs.
The identity operation and the three operations that generate the group are
indicated by their mapping of the set of $16$ Boolean functions with $K\!=\!2$ inputs
onto itself. 
The first column depicts the unchanged Boolean functions, or equivalently the identity operation $I$. 
The other columns show the effect of the symmetry operations rotation $R$, pruning $N$, and parity
$P$ on each of the functions, respectively.
Functions are listed in the first column consistent with the groupings in Fig.~\ref{Figfunction3table}.}
\label{Figgenerators}
\end{figure}

\begin{figure}\begin{center}
\includegraphics[width=8.6cm,keepaspectratio]{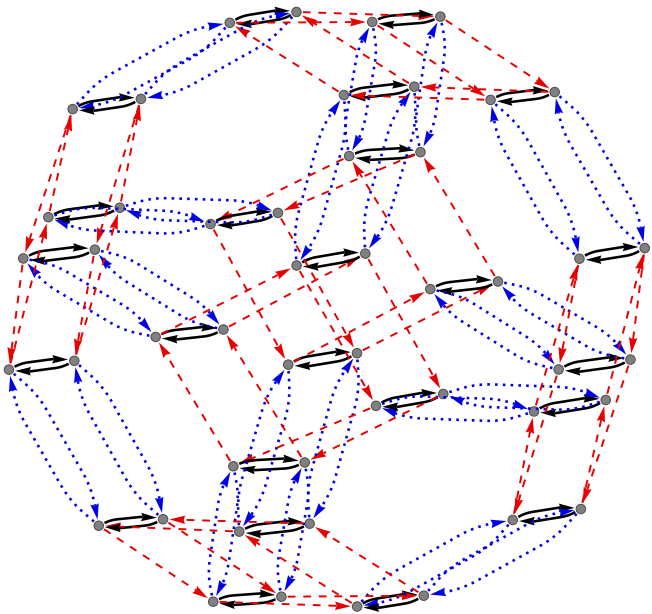}
\end{center}
\vspace{-0.5cm}
\caption{Cayley diagram of the minimal canalization preserving symmetry group for Boolean functions with $K\!=\!2$
inputs.
The elements of the group are the nodes of the graph, shown as gray dots. 
Directed edges of the graph indicate the effect of combining one of the group generators with
an element.
Arrows with different colors correspond to the effect of the 3 different generators 
of the group; rotation $R$, pruning $N$, and parity $P$ operations are shown as red dashed, blue dotted, and black solid 
arrows, respectively.}
\label{Figcayley}
\end{figure}

For Boolean functions with $K\!=\!2$ inputs,
the smallest symmetry group we have identified that has the correct function
classes has $48$ symmetry operations.
This group can be generated by three
symmetry operations that we designate as $R$, $N$ and $P$.
These are shown in Fig.~\ref{Figgenerators}.
$R$ is counterclockwise rotation by $90$
degrees in the geometrical representation of the functions, 
$P$ is the parity operation that maps output 0 into output 1 and
vice versa, 
and $N$ is a \textit{pruning} operation that maps
functions onto one another that have different internal homogeneity but the
the same canalization properties.
The presentation of the group generated by the 
relations between these symmetry operations is
\begin{eqnarray*}
&\begin{array}{lcl}
\{R^{4}=N^{2}=P^{2}=I, (RN)^{3}=I, RP=PR, NP=PN\},
\end{array}
\end{eqnarray*}
where $I$ is the identity operation.
The structure of this group can be seen in the
Cayley diagram shown in Fig.~\ref{Figcayley}. 
In this diagram, 
nodes represent the elements of the group, 
and directed edges of the graph connecting them represent the effects
of the generators of the group.
That is, an edge, $E$, from element $e_{1}$ to element $e_{2}$ shows that $E\cdot e_{1}\!=\!e_{2}$.
In this case, edges with different colors correspond to the effects
of the different generators of the group. 
If only rotations are considered, which are
indicated by the red arrows, the diagram consists of 12 disconnected squares.
Combining parity, shown by the black arrows, with rotation pairs the 12 squares into 
6 disconnected cubes. Finally, the cubes become interconnected with the pruning operation, shown by blue arrows.

Function classes, or orbits, do not, however, uniquely determine a
symmetry group. Indeed, one can find more than one symmetry group 
for each value of in-degree $K$ that preserves canalization. 
This is because the critical state may have additional symmetries, 
which will be reflected in the complete symmetry group of the dynamics. 
The actual symmetry of the critical dynamics of networks with 
$K=2$ may, thus, be larger than the minimal group depicted
in Fig.~\ref{Figgenerators}.
Ideally, one would like to find the smallest group that describes
all symmetries of the critical dynamics of networks for each value
of $K$. This remains a difficult, open question. Nevertheless, the
orbits manifested by a dynamics can be used to distinguish it from 
other dynamics.

\section{Symmetry in Other Dynamics}

The fact that the various Boolean functions behave symmetrically and 
can be grouped together into classes that characterize the symmetry is not
restricted to critical random Boolean networks dynamics. 
The symmetries of other types of dynamics of Boolean networks
can be manifested analogously,
and a similar analysis to that described above can be done to determine 
the symmetry in those cases.
Consider, for example, the adaptive dynamics 
mentioned in the introduction,
in which the
nodes compete in an evolutionary game that causes the network to
self-organize to a critical state~\cite{Paczuski_self-organized_2000}.
This game evolves the output functions used by the nodes through
an extremal process~\cite{bak_punctuated_1993, paczuski_avalanche_1996}
that uses negative reinforcement~\cite{challet_emergence_1997} 
to select against majority behavior~\cite{zhang_modeling_1998}.
One step of the process, 
which occurs on a long time scale referred to as an epoch,
starts with some random initial state of the nodes, 
and finds the dynamical attractor.
It then targets a node for evolution
by finding the one whose output
is most often aligned with the output of the majority of nodes
over the period of the attractor,
and replaces its output function with a randomly chosen Boolean function.
The process continues indefinitely, evolving the output function of
one node each epoch. Over time, the 
set of output functions used by the nodes evolve through the process 
to promote diversity among their dynamics ~\cite{bassler_evolution_2004}.

Through the evolutionary dynamics of the game it has been shown that a network will evolve
to a critical state, evidenced by a power-law distribution
of attractor periods~\cite{Paczuski_self-organized_2000}. 
The evolved state is a steady state, in the sense that ensemble-averaged
properties are time-independent.
For networks of nodes that each have $K$ inputs,
depending on the value of $K$,
this occurs over a wide range of bias $p$
with which the new Boolean functions are randomly
chosen to replace those of the targeted nodes that lose
the game~\cite{reichl_canalization_2011}.
Networks with $K\!=\!3$ inputs per node 
will self-organize to a critical steady state even
when the new Boolean functions are unbiasedly
chosen, that is, chosen with $p\!=\!1/2$.
Note that, as described earlier, the game's evolutionary process
will alter only the nodes' output functions; it does
not explicitly alter the directed edges describing the regulatory
interactions of the nodes. Nevertheless, 
the process does effectively alter the edges,
making the process effectively an adaptive one that
co-evolves both node and edge dynamics~\cite{gross_epidemic_2006}.
This is because some inputs of the output functions can be rendered effectively irrelevant through canalization.
This makes their corresponding input edges effectively irrelevant, and thus, as the output functions evolve,
so will the directed graph that describes the
effective regulatory interactions~\cite{reichhardt_canalization_2007}.

An analysis along similar lines to that described above for finding 
symmetry in the critical state dynamics can also be done
to find symmetry in the adaptive dynamics of the evolutionary
game. In order to do so, however, the distribution of output functions
of all of the nodes should be studied. 
This contrasts with our analysis of critical state dynamics of random Boolean networks, where the distribution of output
functions of only a subset of the nodes, active nodes, was studied. 
This difference is due to the fact that the adaptive dynamics of the game directly affects
the distribution of all the output functions, whereas critical
dynamics involves only a subset of nodes.
Indeed, if an analysis 
of the distribution of the output functions of nodes 
that are active on attractors, 
similar to what we have done for random Boolean networks constructed to be critical, 
is done on the networks evolved to a critical state by the game,
then canalization preserving symmetry, not Zyklenzeiger symmetry,
will be found.

\begin{figure}[t!]\begin{center}
\includegraphics[width=8.6cm,keepaspectratio]{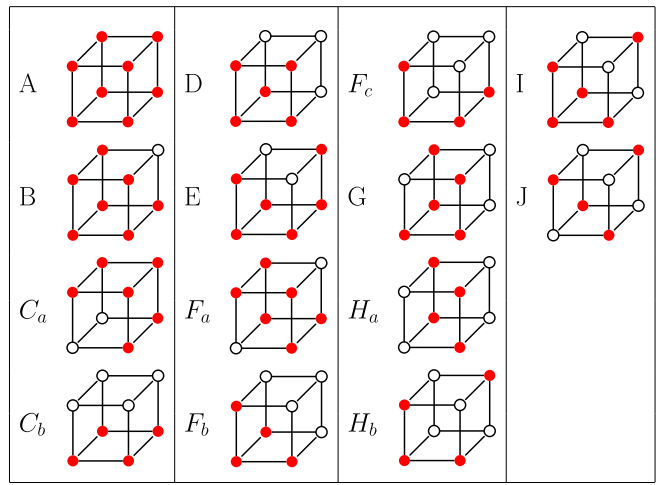}
\end{center}
\vspace{-0.5cm}
\caption{Ising hypercube representation of Boolean functions with $K\!=\!3$ inputs. 
The representation of an example function from each of the $14$ Zyklenzeiger 
classes is shown. 
These classes are labeled in accordance with Ref.~\cite{bassler_evolution_2004}.}
\label{Figk3functions}
\end{figure}

For large networks of $K\!=\!3$ inputs, it has previously been shown that
the ensemble-averaged distribution of output functions used by all
of the nodes in the evolved steady state of the game has 14 different
classes, not the 10 found for critical state dynamics~\cite{reichhardt_canalization_2007}. Thus, the 
symmetry of the adaptive dynamics of the evolutionary game differs
from that of the critical state dynamics. 
In the case of the game, the symmetry corresponds to that of
the $K$-dimensional Zyklenzeiger group,
which consists of octahedral symmetry operations 
combined with the parity operation~\cite{reichhardt_canalization_2007}. 
The 3-dimensional
octahedral symmetry operations are $24$ proper and $24$ improper rotations of the Ising
hypercube~\cite{hargittai_symmetry_2009}, 
which together with parity operator, $P$, make the Zyklenzeiger group a group of order $96$~\cite{harrison_number_1963}. 
The orbits of this group correspond to the $14$ classes of Boolean functions. 
The functions that are in the same class are mapped into each other by the
elements of the Zyklenzeiger group.
The hypercube representation of one representative function from each class is
shown in Fig.~\ref{Figk3functions}.
All of the rest of the 256 functions can be obtained from these representative
functions through the Zyklenzeiger operations.
These 14 Zyklenzeiger classes are labeled from $A$ to $J$, according to the
canalization properties of the functions in the class. Additionally,
those functions labeled
$C$, $F$, and $H$, 
form multiple classes, designated with subscripts, to distinguish those
with the same canalization properties but different internal homogeneity and parity symmetry~\cite{bassler_evolution_2004}. 
The 10 function classes of critical network dynamics result from
combining the Zyklenzeiger classes designated with subscripts.
Thus, some of the canalization preserving classes split to
make the Zyklenzeiger classes.
This suggests that the actual canalization preserving symmetry is
a higher symmetry than the Zyklenzeiger group, and therefore that
the critical dynamics of Boolean networks have a higher symmetry
than does the adaptive evolutionary game.

Still there are other dynamical symmetries for Boolean networks.
In the case of large networks of $K\!=\!2$ inputs, it has been shown that the 2-dimensional 
Zyklenzeiger group has 4 orbits that correspond to the 4 classes of output functions of the 
evolved steady state of the game~\cite{reichhardt_canalization_2007}. 
This behavior is different from that of the critical state dynamics of 
random Boolean networks with 2 inputs that we saw earlier, 
which has 3 orbits,
reflecting the distinct dynamics of these systems. 
Also, if the evolutionary game is played on finite size networks, 
the Zyklenzeiger symmetry is broken, 
and the evolutionary dynamics is controlled 
by the \textit{input-inversion} subgroup of the Zyklenzeiger group~\cite{liu_finite_2011}. 
This is because finite size effects force the nodes to co-operate differently, 
and hence change the underlying dynamics. 
This subgroup has 46 classes for functions with $K\!=\!3$ inputs.
Another example of a different dynamics on Boolean networks is threshold Boolean networks, 
for which the output values of nodes of the network change depending on the comparison 
of the sum of their input values and a threshold value. 
These networks can be used to model genetic regulatory and 
neural networks~\cite{rybarsch_binary_2012, zanudo_boolean_2011}. 
Since threshold networks have a different dynamics from the evolutionary game 
and the random Boolean networks, 
one can expect to find a different symmetry that controls their dynamics,
but the number of orbits that results from this dynamics is not yet known.

Here we see that different dynamics are controlled by distinct symmetries of the critical state, 
and as a result distinct symmetry groups. The characteristic features of the functions that divide 
them into classes give insight into the behavior and structure of the networks. 
For instance, the fact that canalization properties are the characteristic features that classify 
the functions of critical random Boolean networks tells us how the active nodes  
form a substructure of the network. 
Moreover, if one knows the symmetry of the critical state \textit{a priori}, 
it makes the analytical description of the system much easier, 
because in such a description each class of functions can be reduced 
to a single entity. This indeed shows that using symmetry can be a powerful method for analysis 
of the behavior of critical Boolean networks.


\section{Summary and Discussion}

In this paper, we have shown that symmetry can be used to obtain fundamental insights into the dynamics
of heterogeneous complex systems. These systems consist of components that
can behave heterogeneously in the system's dynamics.
When the dynamics has a symmetry, that symmetry can be manifested
in the way that the heterogeneous components behave, causing 
classes of different components to behave in a similar, symmetric manner.
By comparing the behavior of the various components,
characteristic features of the symmetry can be discerned.
Different types of dynamics can be manifested differently,
allowing them to be distinguished.
Furthermore, the symmetric behavior of classes of the components
can be exploited to reduce the complexity of analytic models
of their behavior.

To concretely demonstrate how symmetry is manifested in 
and can be used to help understand
the dynamics of heterogeneous complex systems,
we have used critical random Boolean networks as prototypical examples. 
Random Boolean networks are ideally suited for this purpose because
they are relatively simple models that have heterogeneous
structure and their constituent parts have their own dynamics.
This heterogeneity occurs through the differing output functions of
the nodes and through the directed edges defining the regulatory node interactions.
Importantly, there is also a continuous transition in the nature of the dynamics of random Boolean networks. 
Like other condensed matter systems, their critical states
at the continuous transition are scale invariant, 
making symmetry particularly important and apparent.
For this reason, we have chosen to focus on the critical state properties of
random Boolean networks in order to be as clear as possible.

We have shown that the symmetry underlying the critical dynamics
of random Boolean networks is revealed by  
the ensemble-averaged likelihood that nodes with a given output function 
remain active on the dynamical attractors.
Using numerical simulations, we found that the ratio of the number of nodes utilizing a particular Boolean function that are active on dynamical attractors to the number of all nodes utilizing that function depends solely on the function's canalization properties. 
Canalization describes the ability of a function to have its output fully determined by a subset of its inputs, thus rendering the remaining inputs irrelevant.
That is, all functions with the same canalization properties have the same likelihood of being active. 
Thus, there is a symmetry between all Boolean functions with the same canalization properties. 
Using an analytic model of the critical dynamics involving a stochastic
pruning process that is able to accurately reproduce our numerical results, 
we showed that the symmetry between nodes that have output functions 
with the same canalization properties becomes explicit.

Symmetry of Boolean networks was contrasted with the symmetry underlying the
adaptive networks dynamics of a game that governs the evolution of the output 
functions in a way that causes the network to self-organize to
a critical state. It was previously shown that classes of functions exist for this evolved state. For large networks, these classes are orbits of the Zyklenzeiger symmetry group,
which is distinct from the canalization preserving symmetry group.
For finite-size networks, a breaking of the Zyklenzeiger
symmetry is known to occur, which causes the function classes to
split. Other symmetries that may underlie other types of
dynamics of Boolean networks, such as threshold adaptive dynamics,~\cite{liu_emergent_2006}
could manifest themselves through similar methods of analysis
that compare the frequency with which the various heterogeneous
output functions are utilized by the nodes involved in the dynamics. Symmetries of Boolean network dynamics that appear in this way will always reflect point group symmetries, since Boolean functions are discrete and finite.
In more general types of heterogeneous complex systems,
however, symmetries among the heterogeneous components
can also be continuous. That is, if instead of heterogenous Boolean functions, 
heterogeneous continuous functions are assigned to the nodes of the network, 
the underlying symmetry may be reflected by continuous symmetry groups. 
It would be interesting to extend
the ideas and tools developed in this paper to the more
general cases of systems with continuous symmetries. 

Finally, 
we note that in the analysis of the critical dynamics of Boolean networks 
presented here, we appealed to a prior understanding of the dynamical process
in order to explicitly see its symmetry. 
Such prior understanding, however, is not required. 
In fact, our approach to understanding heterogeneous complex systems 
through analyzing the symmetry of the behavior of different components
may find its greatest benefit 
when used in empirical analyses of the behavior of experimental 
complex systems for which there is no understanding at the outset. In these cases, as we discussed, a direct empirical study of the behavior of the
heterogeneous network components on the various dynamical attractors can be made.


\section{Acknowledgements}

We would like to thank Florian Greil and Mark Tomforde for helpful discussions about this work, and Alireza Fatollahi for careful reading of the manuscript.
This work was supported by the NSF through grants DMR-0908286 and DMR-1206839, and by 
the AFOSR and DARPA through grant FA9550-12-1-0405.

\bibliographystyle{apsrev}
\bibliography{bibliography_shabnamhossein_arxiv}

\begin{thebibliography}{58}
\expandafter\ifx\csname natexlab\endcsname\relax\def\natexlab#1{#1}\fi
\expandafter\ifx\csname bibnamefont\endcsname\relax
  \def\bibnamefont#1{#1}\fi
\expandafter\ifx\csname bibfnamefont\endcsname\relax
  \def\bibfnamefont#1{#1}\fi
\expandafter\ifx\csname citenamefont\endcsname\relax
  \def\citenamefont#1{#1}\fi
\expandafter\ifx\csname url\endcsname\relax
  \def\url#1{\texttt{#1}}\fi
\expandafter\ifx\csname urlprefix\endcsname\relax\def\urlprefix{URL }\fi
\providecommand{\bibinfo}[2]{#2}
\providecommand{\eprint}[2][]{\url{#2}}

\bibitem[{\citenamefont{Griffiths}(2005)}]{griffiths_elementary_2005}
\bibinfo{author}{\bibfnamefont{D.}~\bibnamefont{Griffiths}},
  \textit{\bibinfo{title}{Elementary particles}}
  (\bibinfo{publisher}{Wiley-{VCH}}, \bibinfo{address}{Weinheim},
  \bibinfo{year}{2005}).

\bibitem[{\citenamefont{Kadanoff}(2000)}]{kadanoff_statistical_2000}
\bibinfo{author}{\bibfnamefont{L.~P.} \bibnamefont{Kadanoff}},
  \textit{\bibinfo{title}{Statistical physics: statics, dynamics and
  renormalization}} (\bibinfo{publisher}{World Scientific},
  \bibinfo{address}{Singapore; River Edge, {N.J.}}, \bibinfo{year}{2000}).

\bibitem[{\citenamefont{Boccara}(2010)}]{boccara_modeling_2010}
\bibinfo{author}{\bibfnamefont{N.}~\bibnamefont{Boccara}},
  \textit{\bibinfo{title}{Modeling Complex Systems}}
  (\bibinfo{publisher}{Springer}, \bibinfo{address}{New York, {NY}},
  \bibinfo{year}{2010}).

\bibitem[{\citenamefont{Albert and
  Barab\'{a}si}(2002)}]{albert_statistical_2002}
\bibinfo{author}{\bibfnamefont{R.}~\bibnamefont{Albert}} \bibnamefont{and}
  \bibinfo{author}{\bibfnamefont{A.-L.} \bibnamefont{Barab\'{a}si}},
  \bibinfo{journal}{Reviews of Modern Physics} \textbf{\bibinfo{volume}{74}},
  \bibinfo{pages}{47} (\bibinfo{year}{2002}).

\bibitem[{\citenamefont{Richter and Rost}(2002)}]{richter_komplexe_2002}
\bibinfo{author}{\bibfnamefont{K.}~\bibnamefont{Richter}} \bibnamefont{and}
  \bibinfo{author}{\bibfnamefont{J.-M.} \bibnamefont{Rost}},
  \textit{\bibinfo{title}{Komplexe Systeme}} (\bibinfo{publisher}{Fischer
  Taschenbuch Verlag}, \bibinfo{address}{Frankfurt am Main},
  \bibinfo{year}{2002}).

\bibitem[{\citenamefont{Claudius}(2008)}]{claudius_complex_2008}
\bibinfo{author}{\bibfnamefont{G.}~\bibnamefont{Claudius}},
  \textit{\bibinfo{title}{Complex and adaptive dynamical systems a primer}}
  (\bibinfo{publisher}{Springer}, \bibinfo{address}{Berlin},
  \bibinfo{year}{2008}).

\bibitem[{\citenamefont{Boccaletti et~al.}(2006)\citenamefont{Boccaletti,
  Latora, Moreno, Chavez, and Hwang}}]{boccaletti_complex_2006}
\bibinfo{author}{\bibfnamefont{S.}~\bibnamefont{Boccaletti}},
  \bibinfo{author}{\bibfnamefont{V.}~\bibnamefont{Latora}},
  \bibinfo{author}{\bibfnamefont{Y.}~\bibnamefont{Moreno}},
  \bibinfo{author}{\bibfnamefont{M.}~\bibnamefont{Chavez}}, \bibnamefont{and}
  \bibinfo{author}{\bibfnamefont{D.-U.} \bibnamefont{Hwang}},
  \bibinfo{journal}{Physics Reports} \textbf{\bibinfo{volume}{424}},
  \bibinfo{pages}{175} (\bibinfo{year}{2006}).

\bibitem[{\citenamefont{Derrida and Stauffer}(1986)}]{derrida_phase_1986}
\bibinfo{author}{\bibfnamefont{B.}~\bibnamefont{Derrida}} \bibnamefont{and}
  \bibinfo{author}{\bibfnamefont{D.}~\bibnamefont{Stauffer}},
  \bibinfo{journal}{{EPL} (Europhysics Letters)} \textbf{\bibinfo{volume}{2}},
  \bibinfo{pages}{739} (\bibinfo{year}{1986}).

\bibitem[{\citenamefont{Kauffman}(1969)}]{kauffman_homeostasis_1969}
\bibinfo{author}{\bibfnamefont{S.}~\bibnamefont{Kauffman}},
  \bibinfo{journal}{Nature} \textbf{\bibinfo{volume}{224}},
  \bibinfo{pages}{177} (\bibinfo{year}{1969}).

\bibitem[{\citenamefont{Albert and Othmer}(2003)}]{albert_topology_2003}
\bibinfo{author}{\bibfnamefont{R.}~\bibnamefont{Albert}} \bibnamefont{and}
  \bibinfo{author}{\bibfnamefont{H.~G.} \bibnamefont{Othmer}},
  \bibinfo{journal}{Journal of Theoretical Biology}
  \textbf{\bibinfo{volume}{223}}, \bibinfo{pages}{1} (\bibinfo{year}{2003}).

\bibitem[{\citenamefont{Espinosa-Soto et~al.}(2004)\citenamefont{Espinosa-Soto,
  Padilla-Longoria, and Alvarez-Buylla}}]{espinosa-soto_gene_2004}
\bibinfo{author}{\bibfnamefont{C.}~\bibnamefont{Espinosa-Soto}},
  \bibinfo{author}{\bibfnamefont{P.}~\bibnamefont{Padilla-Longoria}},
  \bibnamefont{and} \bibinfo{author}{\bibfnamefont{E.~R.}
  \bibnamefont{Alvarez-Buylla}}, \bibinfo{journal}{The Plant Cell Online}
  \textbf{\bibinfo{volume}{16}}, \bibinfo{pages}{2923} (\bibinfo{year}{2004}).

\bibitem[{\citenamefont{Pandey et~al.}(2010)\citenamefont{Pandey, Wang, Wilson,
  Li, Zhao, Gookin, Assmann, and Albert}}]{pandey_boolean_2010}
\bibinfo{author}{\bibfnamefont{S.}~\bibnamefont{Pandey}},
  \bibinfo{author}{\bibfnamefont{R.-S.} \bibnamefont{Wang}},
  \bibinfo{author}{\bibfnamefont{L.}~\bibnamefont{Wilson}},
  \bibinfo{author}{\bibfnamefont{S.}~\bibnamefont{Li}},
  \bibinfo{author}{\bibfnamefont{Z.}~\bibnamefont{Zhao}},
  \bibinfo{author}{\bibfnamefont{T.~E.} \bibnamefont{Gookin}},
  \bibinfo{author}{\bibfnamefont{S.~M.} \bibnamefont{Assmann}},
  \bibnamefont{and} \bibinfo{author}{\bibfnamefont{R.}~\bibnamefont{Albert}},
  \bibinfo{journal}{Molecular Systems Biology} \textbf{\bibinfo{volume}{6}}
  (\bibinfo{year}{2010}).

\bibitem[{\citenamefont{Li et~al.}(2004)\citenamefont{Li, Long, Lu, Ouyang, and
  Tang}}]{li_yeast_2004}
\bibinfo{author}{\bibfnamefont{F.}~\bibnamefont{Li}},
  \bibinfo{author}{\bibfnamefont{T.}~\bibnamefont{Long}},
  \bibinfo{author}{\bibfnamefont{Y.}~\bibnamefont{Lu}},
  \bibinfo{author}{\bibfnamefont{Q.}~\bibnamefont{Ouyang}}, \bibnamefont{and}
  \bibinfo{author}{\bibfnamefont{C.}~\bibnamefont{Tang}},
  \bibinfo{journal}{Proceedings of the National Academy of Sciences of the
  United States of America} \textbf{\bibinfo{volume}{101}},
  \bibinfo{pages}{4781} (\bibinfo{year}{2004}).

\bibitem[{\citenamefont{Bornholdt and
  Rohlf}(2000)}]{bornholdt_topological_2000}
\bibinfo{author}{\bibfnamefont{S.}~\bibnamefont{Bornholdt}} \bibnamefont{and}
  \bibinfo{author}{\bibfnamefont{T.}~\bibnamefont{Rohlf}},
  \bibinfo{journal}{Physical Review Letters} \textbf{\bibinfo{volume}{84}},
  \bibinfo{pages}{6114} (\bibinfo{year}{2000}).

\bibitem[{\citenamefont{Wang et~al.}(1990)\citenamefont{Wang, Pichler, and
  Ross}}]{wang_oscillations_1990}
\bibinfo{author}{\bibfnamefont{L.~P.} \bibnamefont{Wang}},
  \bibinfo{author}{\bibfnamefont{E.~E.} \bibnamefont{Pichler}},
  \bibnamefont{and} \bibinfo{author}{\bibfnamefont{J.}~\bibnamefont{Ross}},
  \bibinfo{journal}{Proceedings of the National Academy of Sciences}
  \textbf{\bibinfo{volume}{87}}, \bibinfo{pages}{9467} (\bibinfo{year}{1990}).

\bibitem[{\citenamefont{Aldana et~al.}(2003)\citenamefont{Aldana, Coppersmith,
  and Kadanoff}}]{aldana_boolean_2003}
\bibinfo{author}{\bibfnamefont{M.}~\bibnamefont{Aldana}},
  \bibinfo{author}{\bibfnamefont{S.}~\bibnamefont{Coppersmith}},
  \bibnamefont{and} \bibinfo{author}{\bibfnamefont{L.~P.}
  \bibnamefont{Kadanoff}}, in \textit{\bibinfo{booktitle}{Perspectives and
  Problems in Nolinear Science}}, edited by
  \bibinfo{editor}{\bibfnamefont{E.}~\bibnamefont{Kaplan}},
  \bibinfo{editor}{\bibfnamefont{J.~E.} \bibnamefont{Marsden}},
  \bibnamefont{and} \bibinfo{editor}{\bibfnamefont{K.~R.}
  \bibnamefont{Sreenivasan}} (\bibinfo{publisher}{Springer New York},
  \bibinfo{year}{2003}), pp. \bibinfo{pages}{23--89}.

\bibitem[{\citenamefont{Wang et~al.}(2012)\citenamefont{Wang, Saadatpour, and
  Albert}}]{wang_boolean_2012}
\bibinfo{author}{\bibfnamefont{R.-S.} \bibnamefont{Wang}},
  \bibinfo{author}{\bibfnamefont{A.}~\bibnamefont{Saadatpour}},
  \bibnamefont{and} \bibinfo{author}{\bibfnamefont{R.}~\bibnamefont{Albert}},
  \bibinfo{journal}{Physical Biology} \textbf{\bibinfo{volume}{9}},
  \bibinfo{pages}{055001} (\bibinfo{year}{2012}).

\bibitem[{\citenamefont{Bornholdt}(2005)}]{bornholdt_less_2005}
\bibinfo{author}{\bibfnamefont{S.}~\bibnamefont{Bornholdt}},
  \bibinfo{journal}{Science} \textbf{\bibinfo{volume}{310}},
  \bibinfo{pages}{449} (\bibinfo{year}{2005}).

\bibitem[{\citenamefont{Harrison}(1965)}]{harrison_introduction_1965}
\bibinfo{author}{\bibfnamefont{M.~A.} \bibnamefont{Harrison}},
  \textit{\bibinfo{title}{Introduction to Switching Theory and Automata}}
  (\bibinfo{publisher}{{McGraw-Hill}}, \bibinfo{year}{1965}),
  \bibinfo{edition}{1st} ed., ISBN \bibinfo{isbn}{0070268509}.

\bibitem[{\citenamefont{Green et~al.}(2007)\citenamefont{Green, Leishman, and
  Sadedin}}]{green_emergence_2007}
\bibinfo{author}{\bibfnamefont{D.}~\bibnamefont{Green}},
  \bibinfo{author}{\bibfnamefont{T.}~\bibnamefont{Leishman}}, \bibnamefont{and}
  \bibinfo{author}{\bibfnamefont{S.}~\bibnamefont{Sadedin}}, in
  \textit{\bibinfo{booktitle}{{IEEE} Symposium on Artificial Life, 2007. {ALIFE}
  '07}} (\bibinfo{year}{2007}), pp. \bibinfo{pages}{402--408}.

\bibitem[{\citenamefont{Gong and Socolar}(2012)}]{gong_quantifying_2012}
\bibinfo{author}{\bibfnamefont{X.}~\bibnamefont{Gong}} \bibnamefont{and}
  \bibinfo{author}{\bibfnamefont{J.~E.~S.} \bibnamefont{Socolar}},
  \bibinfo{journal}{Physical Review E} \textbf{\bibinfo{volume}{85}},
  \bibinfo{pages}{066107} (\bibinfo{year}{2012}).

\bibitem[{\citenamefont{Bastolla and Parisi}(1997)}]{bastolla_numerical_1997}
\bibinfo{author}{\bibfnamefont{U.}~\bibnamefont{Bastolla}} \bibnamefont{and}
  \bibinfo{author}{\bibfnamefont{G.}~\bibnamefont{Parisi}},
  \bibinfo{journal}{Journal of Theoretical Biology}
  \textbf{\bibinfo{volume}{187}}, \bibinfo{pages}{117} (\bibinfo{year}{1997}).

\bibitem[{\citenamefont{Bhattacharjya and
  Liang}(1996)}]{bhattacharjya_power-law_1996}
\bibinfo{author}{\bibfnamefont{A.}~\bibnamefont{Bhattacharjya}}
  \bibnamefont{and} \bibinfo{author}{\bibfnamefont{S.}~\bibnamefont{Liang}},
  \bibinfo{journal}{Physical Review Letters} \textbf{\bibinfo{volume}{77}},
  \bibinfo{pages}{1644} (\bibinfo{year}{1996}).

\bibitem[{\citenamefont{Yeomans}(1992)}]{yeomans_statistical_1992}
\bibinfo{author}{\bibfnamefont{J.~M.} \bibnamefont{Yeomans}},
  \textit{\bibinfo{title}{Statistical mechanics of phase transitions}}
  (\bibinfo{publisher}{Clarendon Press ; Oxford University Press},
  \bibinfo{address}{Oxford [England]; New York}, \bibinfo{year}{1992}).

\bibitem[{\citenamefont{Packard}(1988)}]{packard_adaptation_1988}
\bibinfo{author}{\bibfnamefont{N.~H.} \bibnamefont{Packard}}, in
  \textit{\bibinfo{booktitle}{Dynamic Patterns in Complex Systems}}, edited by
  \bibinfo{editor}{\bibfnamefont{J.~A.~S.} \bibnamefont{Kelso}},
  \bibinfo{editor}{\bibfnamefont{A.~J.} \bibnamefont{Mandell}},
  \bibnamefont{and} \bibinfo{editor}{\bibfnamefont{M.~F.}
  \bibnamefont{Shlesinger}} (\bibinfo{publisher}{World Scientic},
  \bibinfo{address}{Singapore}, \bibinfo{year}{1988}), pp.
  \bibinfo{pages}{293--301}.

\bibitem[{\citenamefont{Derrida and Pomeau}(1986)}]{derrida_random_1986}
\bibinfo{author}{\bibfnamefont{B.}~\bibnamefont{Derrida}} \bibnamefont{and}
  \bibinfo{author}{\bibfnamefont{Y.}~\bibnamefont{Pomeau}},
  \bibinfo{journal}{{EPL} (Europhysics Letters)} \textbf{\bibinfo{volume}{1}},
  \bibinfo{pages}{45} (\bibinfo{year}{1986}).

\bibitem[{\citenamefont{Paczuski et~al.}(2000)\citenamefont{Paczuski, Bassler,
  and Corral}}]{Paczuski_self-organized_2000}
\bibinfo{author}{\bibfnamefont{M.}~\bibnamefont{Paczuski}},
  \bibinfo{author}{\bibfnamefont{K.~E.} \bibnamefont{Bassler}},
  \bibnamefont{and} \bibinfo{author}{\bibfnamefont{A.}~\bibnamefont{Corral}},
  \bibinfo{journal}{Physical Review Letters} \textbf{\bibinfo{volume}{84}},
  \bibinfo{pages}{3185} (\bibinfo{year}{2000}).

\bibitem[{\citenamefont{{MacArthur} et~al.}(2009)\citenamefont{{MacArthur},
  S\'{a}nchez-Garc\'{i}a, and Ma'ayan}}]{macarthur_spectral_2009}
\bibinfo{author}{\bibfnamefont{B.~D.} \bibnamefont{{MacArthur}}},
  \bibinfo{author}{\bibfnamefont{R.~J.} \bibnamefont{S\'{a}nchez-Garc\'{i}a}},
  \bibnamefont{and} \bibinfo{author}{\bibfnamefont{A.}~\bibnamefont{Ma'ayan}},
  \bibinfo{journal}{Physical Review E} \textbf{\bibinfo{volume}{80}},
  \bibinfo{pages}{026117} (\bibinfo{year}{2009}).


\bibitem[{\citenamefont{Kauffman}(1993)}]{kauffman_origins_1993}
\bibinfo{author}{\bibfnamefont{S.~A.} \bibnamefont{Kauffman}},
  \textit{\bibinfo{title}{The origins of order: self-organization and selection
  in evolution}} (\bibinfo{publisher}{Oxford University Press},
  \bibinfo{address}{New York}, \bibinfo{year}{1993}).

\bibitem[{\citenamefont{Waddington}(1942)}]{waddington_canalization_1942}
\bibinfo{author}{\bibfnamefont{C.~H.} \bibnamefont{Waddington}},
  \bibinfo{journal}{Nature} \textbf{\bibinfo{volume}{150}},
  \bibinfo{pages}{563} (\bibinfo{year}{1942}).

\bibitem[{\citenamefont{Liu and Bassler}(2011)}]{liu_finite_2011}
\bibinfo{author}{\bibfnamefont{M.}~\bibnamefont{Liu}} \bibnamefont{and}
  \bibinfo{author}{\bibfnamefont{K.~E.} \bibnamefont{Bassler}},
  \bibinfo{journal}{Journal of Physics A: Mathematical and Theoretical}
  \textbf{\bibinfo{volume}{44}}, \bibinfo{pages}{045101}
  (\bibinfo{year}{2011}).

\bibitem[{\citenamefont{Walker and Ashby}(1966)}]{walker_temporal_1966}
\bibinfo{author}{\bibfnamefont{C.~C.} \bibnamefont{Walker}} \bibnamefont{and}
  \bibinfo{author}{\bibfnamefont{W.~R.} \bibnamefont{Ashby}},
  \bibinfo{journal}{Kybernetik} \textbf{\bibinfo{volume}{3}},
  \bibinfo{pages}{100} (\bibinfo{year}{1966}).

\bibitem[{\citenamefont{Bastolla and
  Parisi}(1998{\natexlab{a}})}]{bastolla_relevant_1998}
\bibinfo{author}{\bibfnamefont{U.}~\bibnamefont{Bastolla}} \bibnamefont{and}
  \bibinfo{author}{\bibfnamefont{G.}~\bibnamefont{Parisi}},
  \bibinfo{journal}{Physica D: Nonlinear Phenomena}
  \textbf{\bibinfo{volume}{115}}, \bibinfo{pages}{203}
  (\bibinfo{year}{1998}{\natexlab{a}}).

\bibitem[{\citenamefont{Socolar and Kauffman}(2003)}]{socolar_scaling_2003}
\bibinfo{author}{\bibfnamefont{J.~E.~S.} \bibnamefont{Socolar}}
  \bibnamefont{and} \bibinfo{author}{\bibfnamefont{S.~A.}
  \bibnamefont{Kauffman}}, \bibinfo{journal}{Physical Review Letters}
  \textbf{\bibinfo{volume}{90}}, \bibinfo{pages}{068702}
  (\bibinfo{year}{2003}).

\bibitem[{\citenamefont{Kaufman and Drossel}(2006)}]{kaufman_relevant_2006}
\bibinfo{author}{\bibfnamefont{V.}~\bibnamefont{Kaufman}} \bibnamefont{and}
  \bibinfo{author}{\bibfnamefont{B.}~\bibnamefont{Drossel}},
  \bibinfo{journal}{New Journal of Physics} \textbf{\bibinfo{volume}{8}},
  \bibinfo{pages}{228} (\bibinfo{year}{2006}).

\bibitem[{\citenamefont{Samuelsson and
  Socolar}(2006)}]{samuelsson_exhaustive_2006}
\bibinfo{author}{\bibfnamefont{B.}~\bibnamefont{Samuelsson}} \bibnamefont{and}
  \bibinfo{author}{\bibfnamefont{J.~E.~S.} \bibnamefont{Socolar}},
  \bibinfo{journal}{Physical Review E} \textbf{\bibinfo{volume}{74}},
  \bibinfo{pages}{036113} (\bibinfo{year}{2006}).

\bibitem[{\citenamefont{Kaufman et~al.}(2005)\citenamefont{Kaufman, Mihaljev,
  and Drossel}}]{kaufman_scaling_2005}
\bibinfo{author}{\bibfnamefont{V.}~\bibnamefont{Kaufman}},
  \bibinfo{author}{\bibfnamefont{T.}~\bibnamefont{Mihaljev}}, \bibnamefont{and}
  \bibinfo{author}{\bibfnamefont{B.}~\bibnamefont{Drossel}},
  \bibinfo{journal}{Physical Review E} \textbf{\bibinfo{volume}{72}},
  \bibinfo{pages}{046124} (\bibinfo{year}{2005}).

\bibitem[{\citenamefont{Mihaljev and Drossel}(2006)}]{mihaljev_scaling_2006}
\bibinfo{author}{\bibfnamefont{T.}~\bibnamefont{Mihaljev}} \bibnamefont{and}
  \bibinfo{author}{\bibfnamefont{B.}~\bibnamefont{Drossel}},
  \bibinfo{journal}{Physical Review E} \textbf{\bibinfo{volume}{74}},
  \bibinfo{pages}{046101} (\bibinfo{year}{2006}).

\bibitem[{\citenamefont{Reichhardt and
  Bassler}(2007)}]{reichhardt_canalization_2007}
\bibinfo{author}{\bibfnamefont{C.~J.~O.} \bibnamefont{Reichhardt}}
  \bibnamefont{and} \bibinfo{author}{\bibfnamefont{K.~E.}
  \bibnamefont{Bassler}}, \bibinfo{journal}{Journal of Physics A: Mathematical
  and Theoretical} \textbf{\bibinfo{volume}{40}}, \bibinfo{pages}{4339}
  (\bibinfo{year}{2007}).

\bibitem[{\citenamefont{Schwabl}(2006)}]{schwabl_statistical_2006}
\bibinfo{author}{\bibfnamefont{F.}~\bibnamefont{Schwabl}},
  \textit{\bibinfo{title}{Statistical mechanics}} (\bibinfo{publisher}{Springer},
  \bibinfo{address}{Berlin; New York}, \bibinfo{year}{2006}).

\bibitem[{\citenamefont{Samuelsson and
  Troein}(2003)}]{samuelsson_superpolynomial_2003}
\bibinfo{author}{\bibfnamefont{B.}~\bibnamefont{Samuelsson}} \bibnamefont{and}
  \bibinfo{author}{\bibfnamefont{C.}~\bibnamefont{Troein}},
  \bibinfo{journal}{Physical Review Letters} \textbf{\bibinfo{volume}{90}},
  \bibinfo{pages}{098701} (\bibinfo{year}{2003}).

\bibitem[{\citenamefont{Bastolla and
  Parisi}(1998{\natexlab{b}})}]{bastolla_modular_1998}
\bibinfo{author}{\bibfnamefont{U.}~\bibnamefont{Bastolla}} \bibnamefont{and}
  \bibinfo{author}{\bibfnamefont{G.}~\bibnamefont{Parisi}},
  \bibinfo{journal}{Physica D: Nonlinear Phenomena}
  \textbf{\bibinfo{volume}{115}}, \bibinfo{pages}{219}
  (\bibinfo{year}{1998}{\natexlab{b}}).

\bibitem[{\citenamefont{Bilke and Sjunnesson}(2001)}]{bilke_stability_2001}
\bibinfo{author}{\bibfnamefont{S.}~\bibnamefont{Bilke}} \bibnamefont{and}
  \bibinfo{author}{\bibfnamefont{F.}~\bibnamefont{Sjunnesson}},
  \bibinfo{journal}{Physical Review E} \textbf{\bibinfo{volume}{65}},
  \bibinfo{pages}{016129} (\bibinfo{year}{2001}).

\bibitem[{\citenamefont{M\"{o}ller and Drossel}(2012)}]{moller_formation_2012}
\bibinfo{author}{\bibfnamefont{M.}~\bibnamefont{M\"{o}ller}} \bibnamefont{and}
  \bibinfo{author}{\bibfnamefont{B.}~\bibnamefont{Drossel}},
  \bibinfo{journal}{New Journal of Physics} \textbf{\bibinfo{volume}{14}},
  \bibinfo{pages}{023051} (\bibinfo{year}{2012}).

\bibitem[{\citenamefont{Paul et~al.}(2006)\citenamefont{Paul, Kaufman, and
  Drossel}}]{paul_properties_2006}
\bibinfo{author}{\bibfnamefont{U.}~\bibnamefont{Paul}},
  \bibinfo{author}{\bibfnamefont{V.}~\bibnamefont{Kaufman}}, \bibnamefont{and}
  \bibinfo{author}{\bibfnamefont{B.}~\bibnamefont{Drossel}},
  \bibinfo{journal}{Physical Review E} \textbf{\bibinfo{volume}{73}},
  \bibinfo{pages}{026118} (\bibinfo{year}{2006}).

\bibitem[{\citenamefont{Wagner}(2007)}]{wagner_robustness_2007}
\bibinfo{author}{\bibfnamefont{A.}~\bibnamefont{Wagner}},
  \textit{\bibinfo{title}{Robustness and evolvability in living systems}}
  (\bibinfo{publisher}{Princeton University Press},
  \bibinfo{address}{Princeton, {N.J.;} Woodstock}, \bibinfo{year}{2007}).

\bibitem[{\citenamefont{Bak and Sneppen}(1993)}]{bak_punctuated_1993}
\bibinfo{author}{\bibfnamefont{P.}~\bibnamefont{Bak}} \bibnamefont{and}
  \bibinfo{author}{\bibfnamefont{K.}~\bibnamefont{Sneppen}},
  \bibinfo{journal}{Physical Review Letters} \textbf{\bibinfo{volume}{71}},
  \bibinfo{pages}{4083} (\bibinfo{year}{1993}).

\bibitem[{\citenamefont{Paczuski et~al.}(1996)\citenamefont{Paczuski, Maslov,
  and Bak}}]{paczuski_avalanche_1996}
\bibinfo{author}{\bibfnamefont{M.}~\bibnamefont{Paczuski}},
  \bibinfo{author}{\bibfnamefont{S.}~\bibnamefont{Maslov}}, \bibnamefont{and}
  \bibinfo{author}{\bibfnamefont{P.}~\bibnamefont{Bak}},
  \bibinfo{journal}{Physical Review E} \textbf{\bibinfo{volume}{53}},
  \bibinfo{pages}{414} (\bibinfo{year}{1996}).

\bibitem[{\citenamefont{Challet and Zhang}(1997)}]{challet_emergence_1997}
\bibinfo{author}{\bibfnamefont{D.}~\bibnamefont{Challet}} \bibnamefont{and}
  \bibinfo{author}{\bibfnamefont{Y.-C.} \bibnamefont{Zhang}},
  \bibinfo{journal}{Physica A: Statistical Mechanics and its Applications}
  \textbf{\bibinfo{volume}{246}}, \bibinfo{pages}{407} (\bibinfo{year}{1997}).

\bibitem[{\citenamefont{Zhang}(1998)}]{zhang_modeling_1998}
\bibinfo{author}{\bibfnamefont{Y.-C.} \bibnamefont{Zhang}},
  \bibinfo{journal}{Europhys. News} \textbf{\bibinfo{volume}{29}},
  \bibinfo{pages}{51} (\bibinfo{year}{1998}).

\bibitem[{\citenamefont{Bassler et~al.}(2004)\citenamefont{Bassler, Lee, and
  Lee}}]{bassler_evolution_2004}
\bibinfo{author}{\bibfnamefont{K.~E.} \bibnamefont{Bassler}},
  \bibinfo{author}{\bibfnamefont{C.}~\bibnamefont{Lee}}, \bibnamefont{and}
  \bibinfo{author}{\bibfnamefont{Y.}~\bibnamefont{Lee}},
  \bibinfo{journal}{Physical Review Letters} \textbf{\bibinfo{volume}{93}},
  \bibinfo{pages}{038101} (\bibinfo{year}{2004}).

\bibitem[{\citenamefont{Reichl and Bassler}(2011)}]{reichl_canalization_2011}
\bibinfo{author}{\bibfnamefont{M.~D.} \bibnamefont{Reichl}} \bibnamefont{and}
  \bibinfo{author}{\bibfnamefont{K.~E.} \bibnamefont{Bassler}},
  \bibinfo{journal}{Physical Review E} \textbf{\bibinfo{volume}{84}},
  \bibinfo{pages}{056103} (\bibinfo{year}{2011}).

\bibitem[{\citenamefont{Gross et~al.}(2006)\citenamefont{Gross, {D'Lima}, and
  Blasius}}]{gross_epidemic_2006}
\bibinfo{author}{\bibfnamefont{T.}~\bibnamefont{Gross}},
  \bibinfo{author}{\bibfnamefont{C.~J.~D.} \bibnamefont{{D'Lima}}},
  \bibnamefont{and} \bibinfo{author}{\bibfnamefont{B.}~\bibnamefont{Blasius}},
  \bibinfo{journal}{Physical Review Letters} \textbf{\bibinfo{volume}{96}},
  \bibinfo{pages}{208701} (\bibinfo{year}{2006}).

\bibitem[{\citenamefont{Hargittai}(2009)}]{hargittai_symmetry_2009}
\bibinfo{author}{\bibfnamefont{M.}~\bibnamefont{Hargittai}},
  \textit{\bibinfo{title}{Symmetry Through the Eyes of a Chemist}}
  (\bibinfo{publisher}{Springer}, \bibinfo{year}{2009}), \bibinfo{edition}{3rd}
  ed., ISBN \bibinfo{isbn}{9048118387}.

\bibitem[{\citenamefont{Harrison}(1963)}]{harrison_number_1963}
\bibinfo{author}{\bibfnamefont{M.~A.} \bibnamefont{Harrison}},
  \bibinfo{journal}{Journal of the Society for Industrial and Applied
  Mathematics} \textbf{\bibinfo{volume}{11}}, \bibinfo{pages}{806}
  (\bibinfo{year}{1963}).

\bibitem[{\citenamefont{Rybarsch and Bornholdt}(2012)}]{rybarsch_binary_2012}
\bibinfo{author}{\bibfnamefont{M.}~\bibnamefont{Rybarsch}} \bibnamefont{and}
  \bibinfo{author}{\bibfnamefont{S.}~\bibnamefont{Bornholdt}},
  \bibinfo{journal}{Physical Review E} \textbf{\bibinfo{volume}{86}},
  \bibinfo{pages}{026114} (\bibinfo{year}{2012}).

\bibitem[{\citenamefont{Za\~{n}udo et~al.}(2011)\citenamefont{Za\~{n}udo,
  Aldana, and Mart\'{i}nez-Mekler}}]{zanudo_boolean_2011}
\bibinfo{author}{\bibfnamefont{J.~G.~T.} \bibnamefont{Za\~{n}udo}},
  \bibinfo{author}{\bibfnamefont{M.}~\bibnamefont{Aldana}}, \bibnamefont{and}
  \bibinfo{author}{\bibfnamefont{G.}~\bibnamefont{Mart\'{i}nez-Mekler}}, in
  \textit{\bibinfo{booktitle}{Information Processing and Biological Systems}},
  edited by \bibinfo{editor}{\bibfnamefont{S.}~\bibnamefont{Niiranen}}
  \bibnamefont{and} \bibinfo{editor}{\bibfnamefont{A.}~\bibnamefont{Ribeiro}}
  (\bibinfo{publisher}{Springer Berlin Heidelberg}, \bibinfo{year}{2011}),
  no.~\bibinfo{number}{11} in \bibinfo{series}{Intelligent Systems Reference
  Library}, pp. \bibinfo{pages}{113--151}.

\bibitem[{\citenamefont{Liu and Bassler}(2006)}]{liu_emergent_2006}
\bibinfo{author}{\bibfnamefont{M.}~\bibnamefont{Liu}} \bibnamefont{and}
  \bibinfo{author}{\bibfnamefont{K.~E.} \bibnamefont{Bassler}},
  \bibinfo{journal}{Physical Review E} \textbf{\bibinfo{volume}{74}},
  \bibinfo{pages}{041910} (\bibinfo{year}{2006}).

\end{thebibliography}

\end{document}